\documentclass[11pt]{article}
\usepackage{titlesec}
\titleformat{\paragraph}[runin]{\normalfont\itshape}{\theparagraph.}{.3em}{}[.]\titlespacing{\paragraph}{0pt}{1ex plus .1ex minus .2ex}{.5em}
\usepackage{amsmath,amssymb,bm}
\usepackage[T1]{fontenc}
\usepackage[utf8]{inputenc}
\usepackage{lmodern}
\usepackage{mathtools}
\usepackage{dsfont}
\pdfoutput=1
\usepackage[english]{babel}
\usepackage[letterpaper, hmargin=1in, top=1in, bottom=1.2in, footskip=0.6in]{geometry}
\usepackage{graphicx} 
\usepackage{booktabs} 
\usepackage{color}
\definecolor{aquamarine}{rgb}{0.5, 1.0, 0.83}
\definecolor{ao(english)}{rgb}{0.0, 0.5, 0.0}
\definecolor{armygreen}{rgb}{0.29, 0.33, 0.13}
\definecolor{awesome}{rgb}{1.0, 0.13, 0.32}
\definecolor{ballblue}{rgb}{0.13, 0.67, 0.8}
\definecolor{bittersweet}{rgb}{1.0, 0.44, 0.37}
\definecolor{blue}{rgb}{0.0, 0.0, 1.0}
\definecolor{brinkpink}{rgb}{0.98, 0.38, 0.5}
\definecolor{ballblue}{rgb}{0.13, 0.67, 0.8}
\definecolor{brightturquoise}{rgb}{0.03, 0.91, 0.87}
\definecolor{blue-green}{rgb}{0.0, 0.87, 0.87}
\definecolor{caribbeangreen}{rgb}{0.0, 0.8, 0.6}
\definecolor{cyan}{rgb}{0.0, 1.0, 1.0}
\definecolor{amber(sae/ece)}{rgb}{1.0, 0.49, 0.0}
\graphicspath{ {images/} }
\definecolor{vdarkred}{rgb}{0.6,0,0.2}

\DeclareMathOperator{\tr}{tr}
\usepackage[utf8]{inputenc}
\usepackage{lmodern}

\newcommand*{\deq}{\mathrel{\vcenter{\baselineskip0.65ex \lineskiplimit0pt \hbox{.}\hbox{.}}}=}
\newcommand*{\eqd}{=\mathrel{\vcenter{\baselineskip0.65ex \lineskiplimit0pt \hbox{.}\hbox{.}}}}
\newcommand{\col}{\vcentcolon}
\newcommand{\wick}[1]{{\col\!#1\!\col}}

\definecolor{vdarkred}{rgb}{0.6,0,0.2}
\definecolor{vdarkblue}{rgb}{0,0.2,0.6}
\usepackage[pdftex, colorlinks, linkcolor=vdarkblue,citecolor=vdarkred]{hyperref}

\author{	
J\"urg Fr\"ohlich
\and Antti Knowles
\and Benjamin Schlein
\and Vedran Sohinger
\vspace{0.1cm}\\
}

\title{A path-integral analysis of interacting Bose gases and loop gases}

\begin{document}

\maketitle

\begin{center}
{\large Dedicated to \textit{Joel L. Lebowitz}}
\end{center}

\vspace{1em}

\begin{abstract}
We review some recent results on interacting Bose gases in thermal equilibrium. 
In particular, we study the convergence of the  grand-canonical equilibrium states of 
such gases to their mean-field limits, which are given by the Gibbs measures 
of classical field theories with quartic Hartree-type self-interaction, and to the Gibbs 
states of classical gases of point particles. We discuss various open problems and 
conjectures concerning, e.g., Bose-Einstein condensation, polymers and 
$\vert \boldsymbol{\phi} \vert^{4}$-theory.
\end{abstract}

\section{Description of systems and purpose of analysis}\label{Intro}

In this paper, we study quantum gases of interacting bosons, such as atomic (or molecular) \textit{hydrogen}, \textit{helium} ($^{4}$He), or \textit{rubidium} ($^{85}$Ru). We review several new results arising from a mathematical analysis of such systems and of gases of interacting Brownian paths and loops equivalent to the former. Only formal calculations and arguments are presented, which are not mathematically rigorous, but illuminate nicely the main ideas underlying our analysis. Detailed proofs of our results are contained, or based on methods recently developed, in \cite{FKSS-1}; see also \cite{FKSS-2}. 

The gateway to our study is a functional-integral formulation of the standard quantum-mechanical description of Bose gases as a kind of \textit{scalar imaginary-time field theory}, which can be derived by applying a Hubbard-Stratonovich transformation (see, e.g., \cite {B-K, DI, MZ}) to a well known formal path-integral representation of the partition function and of reduced density matrices of Bose gases; (see, e.g., \cite{CFS, A}, and references given there).  A concise, mathematically rigorous derivation of our representation is given in \cite{FKSS-1}. Alternatively, one can use Ginibre's representation \cite{Ginibre} of Bose gases as statistical systems of interacting Brownian paths and loops to arrive at the same expressions; see \cite{FKSS-2}. In our functional-integral formulation of Bose gases the number, $N$, of species of particles in the Bose gas appears as a parameter that can take arbitrary complex values (see \cite{MZ} for related ideas). We point out simplifications that appear in various limiting regimes corresponding, e.g., to $N\rightarrow \infty$, $N\rightarrow 0$, $M\rightarrow \infty$, where $M$ is the mass of the particles, and we study the mean-field limit of these systems ($M, \rho \rightarrow \infty$, where $\rho$ is the particle density); see \cite{LNR, FKSS-1} and references given there.

The purpose of our analysis can be summarized as follows.
\begin{itemize}
\item{Improve our general understanding of interacting Bose gases in thermal equilibrium: Among many other things, the thermodynamic limit of the Gibbs potential and the reduced density matrices and their properties above and close to the critical temperature of Bose-Einstein condensation (BEC) are studied; see \cite{Ginibre, Su, Ue, FKSS-2}).}
\item{Study the limit of a classical gas of point particles reached when \mbox{$M\rightarrow \infty$} (see, e.g., \cite{FKSS-2}, and references given there).}
\item{Study the mean-field (or classical field) limit of interacting Bose gases, \cite{LNR, FKSS-1}.}
\item{Initiate a rigorous study of the large-$N$ limit and of BEC at large $N<\infty$ (see also \cite {B-K, DI, MZ}).}
\item{Prepare an analysis of the $N\rightarrow 0$ limit of Bose gases, which corresponds to a regularized version of the self-avoiding walk model of polymer chains (see \cite{de Gennes, DP}).}
\item{Present novel evidence for the conjecture that the $\lambda \vert \boldsymbol{\phi} \vert^{4}$-theory 
of an $N$-component complex scalar field $\boldsymbol{\phi}$ has a non-interacting (free) ultraviolet limit in dimension $d\geq 4$; see \cite{Aizenman,JF,A-DC}.}
\end{itemize}

Next, we introduce the systems studied in this paper more precisely. For simplicity we assume that the particles in the Bose gas are spinless. We suppose that the gas is confined to a container, $\Lambda$, of finite size, which, again for simplicity, we choose to be a cube with sides of length $L$ in physical space $\mathbb{R}^{d}$, with $d=1,2,3$ (and $d\geq4$). The Hilbert space of pure state vectors of $n$ spinless bosons confined to $\Lambda$ is given by
\begin{equation}\label{Hilbert space}
\mathcal{H}_{n} \deq L^{2}(\Lambda, dx)^{\otimes_{s}n},
\end{equation}
where $\otimes_{s}n$ denotes an $n$-fold symmetric tensor product, and $dx$ is the Lebesgue measure on $\mathbb{R}^{d}$. The Hamiltonian of a system of $n$ non-relativistic spinless bosons is the operator 
\begin{equation} \label{Hamiltonian}
H_{n}\deq - \sum_{j=1}^{n} \frac{\Delta_{j}}{2M} + \frac{\lambda}{2} \sum_{i,j=1}^{n} v(x_i - x_j)
\end{equation}
acting on $\mathcal{H}_n$, and (for concreteness) we impose periodic boundary conditions on the Laplacians $\Delta_j$ at the boundary $\partial \Lambda$ of $\Lambda$, $j=1,\dots,n$. In expression (2), we use units such that Planck's constant $\hbar = 1$, $M$ is the mass of a particle, $\lambda\geq 0$ is a coupling constant. Moreover, $v$ is a two-body potential, which we assume to have the following properties: (i) $v$ is even, i.e., $v(x)=v(-x), \forall x \in \Lambda$; (ii) $v$ is continuous in $x$ on the torus $\Lambda$ (in particular, $v(0)<\infty$); (iii) $v$ is of \textit{positive type} (repulsive), meaning that its Fourier transform is non-negative. Sometimes it may also be useful to assume that (iv) $v(x)\geq 0, \forall x \in \Lambda$, but this assumption is somewhat unphysical. Moreover, we may want to assume that, in the limit where $\Lambda \nearrow \mathbb{R}^{d}$, \mbox{(v) $v$ has} integrable decay, in order to be able to construct the thermodynamic limit of various quantities. Under assumptions (ii) and (iii), the operators $H_n$ are densely defined, positive, self-adjoint operators on $\mathcal{H}_n$, for all $n<\infty$.

We are interested in studying the \textit{statistical mechanics} of the systems introduced in \eqref{Hilbert space}, \eqref{Hamiltonian} in thermal equilibrium at some fixed positive temperature, $T$, and positive particle density, $\rho\deq \frac{n}{\vert \Lambda \vert}$. The equilibrium state of the Bose gas in the canonical ensemble at temperature $T$ and density $\rho$ is given by the density matrix 
$P_n$ acting on $\mathcal{H}_n$ given by
\begin{equation}\label{density matrix}
P_n\deq Z_{\Lambda}(\beta, \rho)^{-1} \exp(- \beta H_n), \qquad n\approx \rho\, \vert \Lambda \vert, \quad \beta\deq \frac{1}{k_{B} T}\,,
\end{equation}
where $k_B$ is Boltzmann's constant, $Z_{\Lambda}(\beta, \rho)\deq \mathrm{tr}(e^{-\beta H_{n}})$ is the canonical partition function, and 
$$f_{\Lambda}(\beta, \rho)\deq -\frac{k_{B}T}{\vert \Lambda \vert} \ln Z_{\Lambda}(\beta, \rho)$$
is the free energy density of the system. 

It turns out to be convenient to replace the inverse temperature $\beta$, the mass $M$ of the particles and the coupling constant $\lambda$ by parameters $\nu$ and $\lambda_0$, where
\begin{equation}\label{parameters}
\frac{\beta}{M}\eqd \nu \,\,\text{ and either }\, \lambda =\lambda_0 \text{\,\, or  }\,\, \lambda = \lambda_{0} \nu^{2},
\end{equation}
with $\lambda_0$ an arbitrary but fixed constant. If we choose to vary the density $\rho$ of the gas and the parameters $\nu$ and $\lambda_0$ then the inverse temperature becomes redundant, and we henceforth set $\beta=1$.

For later purposes, we also consider gases of $N=1,2,3, \dots$ different species of spinless bosons, all of mass $M$ and interacting among each other through the same two-body potential $v$. We then set $\lambda = \frac{\lambda_{0}}{N+1}, \text{ or } \lambda = \frac{\lambda_{0}\nu^{2}}{N+1}$, with $\lambda_{0}$ an arbitrary, but fixed constant. Thus, the \textit{parameters} to be chosen are 
\begin{equation}\label{parameters}
 \rho\,,\nu\,, N\,, \lambda_{0}\,.
\end{equation}
We will consider the following \textit{limiting regimes.}
\begin{enumerate}
\item{$\nu\searrow 0$, $\lambda =\lambda_{0}$: classical particle limit.}
\item{$\nu \searrow 0$, $\lambda=\lambda_{0}\nu^{2}$: mean-field (or classical field) limit.}
\item{$N\rightarrow \infty$, $\lambda=\frac{\lambda_{0} \nu^{2}}{N+1}$, $\nu\geq0$: spherical-model (or Berlin-Kac) limit.}
\item{$N\searrow 0$, $\lambda =\frac{\lambda_{0} \nu^{2}}{N+1}$, $\nu\geq 0$: SAW (i.e., self-avoiding-walk, or de Gennes) limit.}
\item{$\Lambda \nearrow \mathbb{R}^{d}$, $\nu>0$: thermodynamic limit.}
\end{enumerate}
We propose to study different approaches to these limiting regimes.

As \textit{long-term goals} we would like to improve our grasp of the following important problems in the statistical mechanics of interacting Bose gases.

\textit{Bose-Einstein condensation} of translation-invariant Bose gases in the thermodynamic limit, for $N$ large enough and for $\nu \geq 0$ (at least in dimension $d\geq4$; see also \cite{MZ}). Applying multi-scale analysis (renormalization-group methods) to the representation of Bose gases studied in this paper and in \cite{FKSS-1} might provide insights and lead to progress in this direction; see also \cite{BFKT}.

\textit{Approach to criticality of the systems with $N\simeq 0$}, in $d\geq4$ dimensions, for $\nu\geq 0$, using, e.g., a suitable variant of the \textit{lace expansion} \cite{BS}, or of supersymmetry methods \cite{BBS-1, BBS-2}. 

\textit{Properties of the $\lambda \vert \boldsymbol{\phi} \vert^{4}_{d}$-Euclidean field theory in dimension $d\geq 4$} 
arising when the two-body potential $v$ approaches a $\delta$-function and $\nu\searrow 0$, with 
$\lambda= \lambda_{0} \nu^{2}$, for $N=1$ (see \cite{JF}); studying Bose gases with $\delta$-function two-body interactions for large values of $N$, and in the de Gennes limit $N\searrow 0$, for arbitrary values of $\nu\geq 0$; see also \cite{BS, BBS-1, BBS-2}.

These are ambitious goals that we have not reached, to date. But we hope that the methods developed in this paper and in \cite{FKSS-1} and \cite{FKSS-2} are useful steps in attempting to approach these goals.

\paragraph{Organization of the paper} 
In the next section, we present a brief review of second quantization and of the grand-canonical ensemble in the equilibrium statistical mechanics of interacting Bose gases. We also identify the classical Hartree field theory that describes the mean-field limit of these gases, which is analyzed in more detail in Section \ref{sec:lg}.
In Section \ref{sec:pi}, we review the (formal) path integral representation of Bose gases in the grand-canonical ensemble. We then re-write the formal functional integrals representing the grand partition function and the Green functions by introducing a Hubbard-Stratonovich transformation. This leads to expressions that are mathematically meaningful. We also use a Feynman-Kac formula to derive some basic bounds on the integrands of those functional integrals and on the grand partition function.
In Section \ref{sec:lg}, we use the results of Section \ref{sec:pi} to derive a representation of interacting Bose gases in terms of interacting Brownian paths and loops, due to Ginibre, and Symanzik's loop gas representation of scalar imaginary-time field theories with quartic self-interactions, which, in our context, appear as mean-field limits of Bose gases. We then describe some basic ideas underlying our proof of convergence to the mean-field limit. In Section \ref{sec:open_problems}, we present some further results that can be derived from the formalism developed in Sections \ref{sec:pi} and \ref{sec:lg}, and we comment on the \textit{limiting regimes} and the \textit{long-term goals} described above. We draw the readers' attention to various important open problems. In Section \ref{sec:ep}, we conclude the paper with a few personal remarks.

\paragraph{Acknowledgements} We thank David Brydges, Alessandro Pizzo and Daniel Ueltschi for very useful discussions and some correspondence on problems related to the ones studied in this paper. We are grateful to Mathieu Lewin, Phan Th\`{a}nh Nam and Nicolas Rougerie for informing us about their beautiful results \cite{LNR} prior to publication.
AK gratefully acknowledges the support of the European Research Council through the RandMat grant and of the Swiss National Science Foundation through the NCCR SwissMAP. BS gratefully acknowledges partial support from the NCCR SwissMAP, from the Swiss National Science Foundation through the Grant ``Dynamical and energetic properties of Bose-Einstein condensates'' and from the European Research Council through the ERC-AdG CLaQS.

\section{The grand-canonical ensemble of interacting Bose gases} \label{sec:gc}
When studying Bose gases in thermal equilibrium, it is often most convenient to use the \textit{grand-canonical ensemble}\footnote{A broad introduction to equilibrium statistical mechanics, including a mathematical discussion of the equivalence of the three standard ensembles -- micro-canonical, canonical and grand-canonical -- can be found in \cite{Ruelle-1}.}: The container 
$\Lambda$ of the gas is connected to a very large reservoir that is kept at a fixed \textit{temperature} and \textit{chemical potential} and contains arbitrarily many particles; as a consequence, the number, $n$, of particles contained in $\Lambda$ is allowed to fluctuate, but its mean value, 
$\langle n \rangle_{\beta, \mu}\equiv \rho \, \vert \Lambda \vert$, is tuned by the chemical potential, 
 $\mu$, of the reservoir. We set 
 $$\beta\, \mu\eqd -  \nu\, \kappa\,, $$
 and, as discussed in the last section, we may henceforth set $\beta\equiv \frac{1}{k_{B}T}=1$ and vary $\nu\propto M^{-1}, \kappa$ and the coupling constant $\lambda$ of the two-body potential.
 
 Next, we briefly recall the formalism of \textit{second quantization}. Let
 \begin{equation}\label{Fock}
 \mathcal{F}_{\Lambda}\deq \bigoplus_{n=0}^{\infty} \mathcal{H}_{n}
 \end{equation}
 be standard \textit{Fock space}, with $\mathcal{H}_n$ as in \eqref{Hilbert space}, and let
$\Phi^{*}_{\nu}(x), \Phi_{\nu}(x)$ denote the usual creation- and annihilation operators (operator-valued distributions) acting on $\mathcal{F}_{\Lambda}$, which satisfy the canonical commutation relations (CCR) 
\begin{equation}\label{CCR}
[\Phi_{\nu}^{\#}(x), \Phi_{\nu}^{\#}(y)]=0, \qquad [\Phi_{\nu}(x), \Phi_{\nu}^{*}(y)]= \nu \cdot\delta(x-y).
\end{equation}
Apparently, the parameter $\nu$ plays the role of Planck's constant $\hbar$, and letting $\nu$ approach $0$ (i.e., $M\rightarrow \infty$) corresponds to approaching a classical limit.
The vector $\Omega\deq\big(1, 0,0,\dots\big) \in \mathcal{F}_{\Lambda}$ is called \textit{vacuum}. One has that
\begin{equation}\label{vacuum}
 \Phi_{\nu}(f)\Omega =0, \,\,\,\,\forall f \in L^{2}(\Lambda, dx),
 \end{equation}
and the set of vectors obtained by applying polynomials in creation operators 
$\Phi_{\nu}^{*}(f), f \in L^{2}(\Lambda, dx)$, to the vacuum $\Omega$ is dense in $\mathcal{F}_{\Lambda}$.
Here 
$$\Phi_{\nu}(f)\deq \int dx\, \overline{f(x)}\, \Phi_{\nu}(x), \quad \Phi_{\nu}^{*}(f)\deq \big(\Phi_{\nu}(f)\big)^{*}.$$
To describe a gas of $N$ species of bosons, one introduces $N$ commuting sets of creation and annihilation operators, 
$\Phi^{*}_{\nu, a}$ and $\Phi_{\nu, a}, a=1,\dots, N$.

Next, we introduce the Hamiltonian describing a Bose gas of $N$ species of bosons
 \begin{equation}\label{gcHam}
  \mathbb{H}_{\nu,\Lambda}\deq\frac{1}{2}\sum_{a=1}^{N}\int_{\Lambda} dx\, \biggl\{ \nabla\Phi_{\nu,a}^{*}(x) \cdot \nabla\Phi_{\nu,a}(x)+ \frac{\lambda}{\nu^{2}}\sum_{b=1}^{N} \int_{\Lambda} dy\, \Phi^{*}_{\nu,a}(x)\Phi_{\nu,a}(x)\, v(x-y) \, \Phi^{*}_{\nu,b}(y) \Phi_{\nu,b}(y)\bigg\}\,.
 \end{equation}
 For $\lambda\geq 0$ and under our assumptions on the two-body potential $v$, this is a well-defined, positive, self-adjoint operator on Fock space $\mathcal{F}_{\Lambda}$.
 We define
 $$\mathbb{N}_{\nu,\Lambda}\deq \sum_{a=1}^{N}\int_{\Lambda} dx \,\Phi_{\nu,a}^{*}(x)\,\Phi_{\nu,a}(x)\,.$$
 The operator $\nu^{-1}\,\mathbb{N}_{\nu,\Lambda}$ counts the number of particles contained in $\Lambda$.
 The \textit{grand partition function} of the gas is defined by
 \begin{equation}\label{grand pf}
 \Xi_{\nu, \Lambda}(\kappa)\deq \mathrm{tr}\Big(\exp[-\mathbb{H}_{\nu,\Lambda} - \kappa \mathbb{N}_{\nu,\Lambda}]\Big)\,,
 \end{equation}
 and the reduced density matrices by
 \begin{equation}\label{RDM}
\gamma_{p}(\underline{x}, \underline{a}; \underline{y}, \underline{a}) \deq \nu^{-p} \,\Xi_{\nu, \Lambda}(\kappa)^{-1} \mathrm{tr}\Big(\exp[- \mathbb{H}_{\nu,\Lambda} - \kappa \mathbb{N}_{\nu, \Lambda} ]\, \Pi_{i=1}^{p} \Phi_{\nu, a_{i}}^{*}(x_i)\,\Pi_{j=1}^{p} \Phi_{\nu,a_{j}}(y_j)\Big)\,,
\end{equation}
where we use the notation that $\underline{x}=(x_1,\dots,x_p)\in \Lambda^{p}$, and $\underline{a}=(a_1,\dots,a_n)$. In the following, it will be convenient to re-write formulae \eqref{grand pf} and \eqref{RDM} by absorbing the chemical potential $\mu$ ($\propto - \kappa$) into a new parameter, denoted (again) by $\rho$, that is used to tune the particle density of an interacting Bose gas, with $\lambda$ strictly positive. We define an operator
\begin{align} \notag
\mathbb{H}_{\Lambda}(\nu, \rho) \deq \frac{1}{2} & \sum_{a=1}^{N} \int_{\Lambda} dx\,\bigg\{\nabla\Phi_{\nu,a}(x)\cdot \nabla \Phi_{\nu,a}(x) + 2\kappa_0 \Phi^{*}_{\nu,a}(x)\Phi_{\nu,a}(x)
\\
\label{quartic}
&+ \lambda \nu^{-2} \sum_{b=1}^{N} \int_{\Lambda} dy\, \big(\Phi^{*}_{\nu, a}(x)\Phi_{\nu, a}(x)-\rho\big)\, v(x-y) \, \big(\Phi^{*}_{\nu,b}(y) \Phi_{\nu,b}(y)-\rho\big)\bigg\},
\end{align}
where $\kappa_0>0$ is an arbitrary constant (in the following kept fixed -- we may set $\kappa_0$ to $1$), and 
$\rho$ is a parameter that can be adjusted to tune the particle density of the interacting gas ($\lambda >0$) to any desired value. When expanding the term on the second line of \eqref{quartic} in $\rho$ we see that, up to a constant, turning on the parameter $\rho$ amounts to adding the quantity $\rho \lambda \nu^{-1} \int_{\Lambda} dx\, v(x)>0$ to the chemical potential, $\mu$, of the Bose gas.
The grand-canonical equilibrium state of the system is then given by the density matrix 
\begin{equation}\label{gc density matrix}
\mathbb{P}_{\Lambda}\deq\Xi_{\Lambda}(\nu,\rho)^{-1} \exp\big[-\mathbb{H}_{\Lambda}(\nu,\rho)\big], \qquad \text{with  }\,\, \Xi_{\Lambda}(\nu,\rho)= \mathrm{tr}\big(\exp[-\mathbb{H}_{\Lambda}(\nu, \rho)]\big)
\end{equation}
acting on Fock space $\mathcal{F}_{\Lambda}$.

It is quite easy to study the quantum dynamics of such Bose gases close to thermal equilibrium, in the sense of constructing and analyzing the time-dependent equilibrium correlations of monomials in $\Phi^{*}_{\nu,a}$ and 
$\Phi_{\nu,a}$, \mbox{$a=1,\dots,N$}; see, e.g., \cite{Ruelle-2}.

Our aim in this paper and in \cite{FKSS-1, FKSS-2} is to study various physical properties of the equilibrium state in \eqref{gc density matrix}, as the parameters $\nu, \lambda$ and $\rho$ are varied. An example is a proof of convergence to the \textit{mean-field (classical field) limit}, which we describe next. 
Recalling that $\nu$ plays the role of Planck's constant and choosing $\lambda = \frac{\lambda_{0}\nu^{2}}{N+1}$ in Eq.~\eqref{gcHam}, we observe that, \textit{formally}, the quantum theory of the Bose gas approaches a
\textit{classical field theory}, as $\nu \searrow 0$, whose Hamilton (energy) functional is given by
\begin{equation}\label{class lim}
\mathfrak{H}(\boldsymbol{\bar{\phi}}, \boldsymbol{\phi})\deq \frac{1}{2} \int_{\Lambda}dx\, \bigg\{\nabla \boldsymbol{\bar{\phi}}(x)\cdot 
\nabla \boldsymbol{\phi}(x) + \frac{\lambda_0}{N+1}\int_{\Lambda} dy\,\vert \boldsymbol{\phi}(x)\vert^{2}\, v(x-y)\, \vert \boldsymbol{\phi}(y)\vert^{2}\bigg\}\,,
\end{equation}
where
$$\boldsymbol{\phi} \deq\begin{pmatrix}\phi_1\\ \vdots\\ \phi_N \end{pmatrix}\quad \text{  and  }\quad \boldsymbol{\bar{\phi}}= 
\big(\bar{\phi}_{1} \dots \bar{\phi}_{N}\big).$$ 
That this so-called mean-field limit makes sense mathematically is a key result discussed in this paper (see Section \ref{sec:open_problems}) and is proven in \cite{FKSS-1} and \cite{LNR}; see \cite{LNR-2, FKSS-3} for earlier results.
The Hamilton functional $\mathfrak{H}$ is well defined on a complex phase space equal to the complex Sobolev space, $\mathcal{H}_1$, over the domain $\Lambda$ equipped with the Poisson brackets
$$\big\{\phi_{a}(x), \bar{\phi}_{b}(y) \big\}=i \delta_{ab}\,\delta(x-y), \qquad \big\{ \phi^{\#}_{a}(x), \phi^{\#}_{b}(y)\big\} =0\,,\quad \forall \,a, b =1,\dots N.$$
The Hamiltonian equations of motion of the classical field theory turn out to be given by the Hartee equation.

Our main interest in this paper is to analyze the convergence properties of the equilibrium states 
$\mathbb{P}_{\Lambda}$ defined in \eqref{gc density matrix}, as $\nu \searrow 0$. It is convenient to define a renormalized Hamilton functional
\begin{align}\notag
\mathfrak{H}(\boldsymbol{\bar{\phi}}, \boldsymbol{\phi};\rho)\deq \frac{1}{2} \int_{\Lambda}dx\, &\bigg\{\nabla \boldsymbol{\bar{\phi}}(x)\cdot 
\nabla \boldsymbol{\phi}(x) + 2\kappa_{0} \vert \boldsymbol{\phi}\vert^{2}
\\
\label{ren Ham}
&+\frac{\lambda_0}{N+1} \int_{\Lambda} dy \, \bigl(\wick{\vert \boldsymbol{\phi}(x)\vert^{2}} -\rho\bigr)\, v(x-y)\, \bigl(\wick{\vert \boldsymbol{\phi}(y)\vert^{2}} -\rho \bigr)\bigg\}\,,
\end{align}
where the double colons $\wick{(\cdot)}$ indicate Wick ordering (with respect to the Gaussian measure on $\mathcal{S}'(\Lambda)$ with mean $0$ and covariance $(-\Delta + 2\kappa_0)^{-1}$). If, in \eqref{quartic}, we replace $\rho$ by $\rho+\delta \rho_{\nu}$, where, in two and three dimensions, $\delta \rho_{\nu}$ must be chosen to diverge at a specific rate, as $\nu\searrow 0$, then the reduced density matrices in the state $\mathbb{P}_{\Lambda}$ defined in 
\eqref{gc density matrix} converge to moments of the \textit{Gibbs measure} of the classical field theory given by
\begin{equation}\label{Gibbs measure}
d\,\mathfrak{P}_{\Lambda}(\boldsymbol{\bar{\phi}}, \boldsymbol{\phi})\deq Z_{\Lambda}^{-1}\, 
\exp[-\mathfrak{H}(\boldsymbol{\bar{\phi}}, \boldsymbol{\phi};\rho)]\, \mathcal{D}\boldsymbol{\bar{\phi}}\wedge \mathcal{D}\boldsymbol{\phi}\,,
\end{equation}
where 
$$\mathcal{D}\boldsymbol{\bar{\phi}}\wedge \mathcal{D}\boldsymbol{\phi} \deq \prod_{x\in \Lambda}\, \prod_{a=1}^{N}d\bar{\phi}_{a}(x) \wedge d\phi_{a}(x)$$ 
is the (\textit{formal}) Liouville measure.
This measure exhibits a continuous $O(N)$-symmetry. A rigorous construction of this measure is an easy task in constructive quantum field theory; see \cite{GJ, Simon}. The proof of convergence, as $\nu\searrow 0$, is, however, \textit{not} easy. It will be sketched in subsequent sections and is the main result established in \cite{FKSS-1, LNR}.

The limiting model, as $N\rightarrow \infty$, is a continuum version of the Berlin-Kac spherical model. In the thermodynamic limit $\Lambda \nearrow \mathbb{R}^{d}$, with $d\geq 3$, this model is known to exhibit a phase transition accompanied by spontaneous symmetry breaking, which is related to BEC. It is an important open problem to prove this for the Gibbs measure introduced in \eqref{Gibbs measure}, for $N<\infty$, in dimension $d\geq3$.

Solving the Hamiltonian equations of motion determined by the Hamilton functional in \eqref{class lim} for initial conditions in the support of the measure $d\,\frak{P}_{\Lambda}$ is fairly easy in one dimension, because a typical sample field configuration $\boldsymbol{\phi}$ in the support of $d\,\mathfrak{P}_{\Lambda}$ is Sobolev regular of index $s<\frac{1}{2}$; see, e.g., \cite{Bourgain-1,FKSS-3}. But, in dimension $d=2$, hard analysis is required to solve this problem, because typical sample configurations in the support of $d\,\mathfrak{P}_{\Lambda}$ are distributional; see \cite{Bourgain-2,Bourgain-3} for relevant results.

Next, we describe a functional-integral construction of the equilibrium state of Bose gases analogous to the Gibbs measure \eqref{Gibbs measure}, but for arbitrary values of the parameter $\nu\geq 0$. The resulting expressions will turn out to be particularly convenient to explore properties of interacting Bose gases under variations of the parameters $\nu, \lambda, \rho$ and $N$ and to study the thermodynamic limit $\Lambda \nearrow \mathbb{R}^{d}$.

\section{Path-integral representation of Bose gases in the grand-canonical ensemble} \label{sec:pi}

\begin{quote}
\textit{A great deal of my work is just playing with equations and seeing what they give.}
\\[0.5em]
\hspace*{\fill} --- P.A.M. Dirac
\end{quote}


In this section, we sketch how path-integral quantization, as originally discovered by \textit{Dirac} in \cite{Dirac}, can be applied to a Bose gas of $N$ species of particles in thermal equilibrium. We \textit{``play with (formal) equations''}; rigorous justifications appear in \cite{FKSS-1}. We propose to express the grand partition function $\Xi_{\Lambda}(\nu,\rho)$ and the reduced density matrices 
$\gamma_{p}$, see \eqref{gc density matrix}, \eqref{RDM}, in terms of functional integrals (see, e.g., \cite{CFS,A}, and references given there). For this purpose, we introduce the \textit{formal} integration measure
$$\mathcal{D}\boldsymbol{\bar{\varphi}} \wedge \mathcal{D}\boldsymbol{\varphi}\deq \prod_{\tau \in [0,\nu)} \prod_{x\in \Lambda}\, \prod_{a=1}^{N}d\bar{\varphi}_{a}(\tau,x) \wedge d\varphi_{a}(\tau,x)\,$$
on the space of configurations of fields $(\bar{\varphi}_{a}, \varphi_a)_{a=1,\dots,N}$ that are periodic in the imaginary-time variable $\tau$ with period $\nu$.
We then find that
\begin{align} \notag
\Xi_{\Lambda}(\nu,\rho) \propto \int  \mathcal{D}\boldsymbol{\bar{\varphi}} \wedge \mathcal{D}\boldsymbol{\varphi} \,\, \exp &\bigg(-\int_{0}^{\nu} d\tau \sum_{a=1}^{N} \int_{\Lambda} dx \bigg\{ 
\bar{\varphi}_{a}(\tau,x) (K_{\Lambda}\varphi_{a})(\tau,x)\\
\label{path integral}
+ \frac{\lambda}{2\nu} \sum_{b=1}^{N} &\int_{\Lambda} dy \big[\vert \varphi_{a}(\tau,x)\vert^{2}-\nu^{-1}\rho\big]\, v(x-y)\, \big[\vert \varphi_{b}(\tau,y)\vert^{2}-\nu^{-1}\rho\big]\bigg\} \bigg)\, 
\end{align}
where
\begin{equation}\label{inv cov}
K_{\Lambda}(\sigma)\deq \frac{\partial}{\partial \tau} - \frac{\Delta}{2} + \kappa_0 + i \sigma(\tau,x), \quad K_{\Lambda}\deq K_{\Lambda}(0)=K_{\Lambda}(\sigma\equiv 0)\,,
\end{equation}
with periodic boundary conditions imposed on $\Delta$ at $\partial \Lambda$.
Reduced density matrices and the so-called Duhamel (imaginary-time) Green functions can be expressed as integrals of monomials in $\big(\bar{\varphi}_{a}(\tau,x), \varphi_{a}(\tau,x)\big), a=1,\dots,N,$ with respect to the \textit{formal} measure given by
\begin{align} 
\notag
\Xi_{\Lambda}(\nu,\rho)^{-1}\,\exp &\Big(-\int_{0}^{\nu} d\tau \sum_{a=1}^{N} \int_{\Lambda} dx \Bigl\{ 
\bar{\varphi}_{a}(\tau,x) (K_{\Lambda}\varphi_{a})(\tau,x) 
\\
\label{complex measure}
&+ \frac{\lambda}{2\nu} \sum_{b=1}^{N} \int_{\Lambda} dy \big[\vert \varphi_{a}(\tau,x)\vert^{2}-\nu^{-1}\rho \big]\, v(x-y)\, \big[\vert \varphi_{b}(\tau,y)\vert^{2}-\nu^{-1}\rho \big]\Big\} \Big)\,\mathcal{D}\boldsymbol{\bar{\varphi}} \wedge \mathcal{D}\boldsymbol{\varphi}\,.
\end{align} 

\paragraph{Remark} In the next section we will discuss the \textit{mean-field limit} of Bose gases. The result of our discussion can be anticipated by studying the behavior of the integrand in \eqref{path integral}, as $\nu$ approaches $0$. Since the fields $\varphi_{a}^{\#}(\tau, x)$ are \textit{periodic} in $\tau$ with period $\nu$, all their Fourier modes corresponding to \textit{non-zero} frequencies ``shoot through the roof'' -- thanks to the presence of the operator 
$\partial/\partial \tau$ in $K_{\Lambda}$ -- and can therefore be neglected, as $\nu \searrow 0$. Thus, for very large values of 
$\nu$, the integrand in \eqref{path integral} is dominated by field configurations 
$$\varphi_{a}^{\#}(\tau, x)\eqd \nu^{-1/2} \phi_{a}^{\#}(x), \qquad a=1,\dots,N,$$
 that are \textit{constant} in $\tau$, and the formal measure introduced in \eqref{complex measure}
approaches the measure
$d\,\mathfrak{P}_{\Lambda}(\boldsymbol{\phi}^{*}, \boldsymbol{\phi})$ defined in \eqref{Gibbs measure}, as $\nu\searrow 0$, provided the parameter $\rho$ in \eqref{path integral} and \eqref{complex measure} is chosen to depend on $\nu$ in a suitable way that amounts to Wick ordering $\vert \varphi_{a}\vert^{2}$ and will be described more precisely, below.

 We emphasize that expression \eqref{complex measure} is \textit{not} a well-defined complex measure. Yet, for 
 $\lambda =0$, i.e., for an ideal Bose gas, one can formally calculate its generating function (done next), which can be used to provide a safe starting point for a mathematically rigorous analysis of interacting Bose gases.

Let $d\mu_{\lambda}(\sigma)$ be the Gaussian probability measure on the space 
$\mathcal{S}'([0,\nu)\times \Lambda)$ of tempered distributions, $\sigma(\tau,x)$, with mean $0$ and a covariance, $C$, given by 
\begin{equation}\label{covariance}
C(\tau,\tau'; x,y)\deq\delta(\tau-\tau')\cdot \frac{\lambda}{\nu}\, v(x-y)\,.
\end{equation}
It then follows from Eq.~\eqref{path integral} by functional Fourier (or Hubbard-Stratonovich) transformation that 
\begin{align}
\Xi_{\Lambda}(\nu,\rho) &= \text{const.} \int  \mathcal{D}\boldsymbol{\bar{\varphi}} \wedge \mathcal{D}\boldsymbol{\varphi} \int d\mu_{\lambda}(\sigma)\,
e^{iN\theta(\sigma)}\, \exp\bigg[-\int_{0}^{\nu} d\tau \int_{\Lambda} dx \, \sum_{a=1}^{N} \bar{\varphi}_{a}(\tau,x)(K_{\Lambda}(\sigma)\varphi_{a})(\tau,x)\bigg] \nonumber \\ \label{Hubbard-Stratonovich}
&= \text{const.} \int d\mu_{\lambda}(\sigma)\,e^{iN\theta(\sigma)}\, [\det\,K_{\Lambda}(\sigma)]^{-N}\,, 
\end{align}
where const.\ stands for a divergent constant, $K_{\Lambda}(\sigma)$ is defined in \eqref{inv cov}, and the phase 
$\theta(\sigma)$ is defined by
\begin{equation}\label{phase}
\theta(\sigma)\deq \frac{\rho}{\nu} \int_{0}^{\nu}d\tau\int_{\Lambda} dx\, \sigma(\tau,x)\,.
\end{equation}
The first equation in \eqref{Hubbard-Stratonovich} follows from the standard formula
\begin{equation}\label{Gauss int}
\int d\mu_{\lambda}(\sigma) \,e^{i \sigma(f)} = \exp[-\langle f, C f \rangle]\,,
\end{equation}
with $\sigma(f)\deq \int_{0}^{\nu} d\tau \int_{\Lambda} dx\, \sigma(\tau,x)\,f(\tau,x)\,.$
The second equation in \eqref{Hubbard-Stratonovich} follows by formally interchanging integration over 
$(\bar{\varphi}_a, \varphi_a)_{a=1,\dots,N}$ with integration over $\sigma$.
For an ideal Bose gas ($\lambda=0$), one formally has that
$$\Xi_{\Lambda}^{(0)}(\nu,\rho) = \text{const.} \,[\det\,K_{\Lambda}(0)]^{-N},$$ 
with the same (divergent) constant as in \eqref{Hubbard-Stratonovich}. It is not hard to see that the \emph{relative partition function}
\begin{equation}\label{relative pf-1}
\Xi_{\Lambda}^{\text{rel}}(\nu,\rho)\deq\frac{\Xi_{\Lambda}(\nu,\rho)}{\Xi_{\Lambda}^{(0)}(\nu)}= \int d\mu_{\lambda}(\sigma)\, e^{iN\theta(\sigma)}\,\bigg[ \frac{\det\, K_{\Lambda}(\sigma)}{\det\,K_{\Lambda}(0)} \bigg]^{-N}
\end{equation}
is actually a mathematically meaningful expression.
Next, we use the simple identity
\begin{equation}\label{det identity}
(\det\,A)^{-1}=\exp\big(-\mathrm{tr} [\ln A]\big)= \exp\bigg(\int_{0}^{\infty}dt\, \mathrm{tr}\big[(A+t)^{-1} - (1+t)^{-1}\big]\bigg)\,,
\end{equation}
for an arbitrary (finite) matrix $A$, with $A+A^{*} >0$, to rewrite \eqref{relative pf-1} as
\begin{equation}\label{relative pf-2}
\Xi_{\Lambda}^{\text{rel}}(\nu,\rho) = \int d\mu_{\lambda}(\sigma)\, e^{iN\theta(\sigma)}\, 
e^{N\,\int_{0}^{\infty}dt\,
\mathrm{tr} \big[(K_{\Lambda}(\sigma)+t)^{-1}- (K_{\Lambda}(0)+t)^{-1}\big]}\,.
\end{equation}
We note that the trace appearing in the exponent on the RHS of \eqref{relative pf-2} is \textit{finite}, and the $t$-integration \textit{converges}, for arbitrary $N<\infty, \rho<\infty$ and $\nu>0$. For $\lambda =\mathcal{O}(\frac{1}{N}),$
one can use \textit{saddle point methods} to calculate the asymptotics of \eqref{relative pf-2}, as $N\rightarrow \infty$. This yields what is known as the \textit{$\frac{1}{N}$-expansion}, which is a powerful tool to study the phase transition  to a high-density phase exhibiting BEC in interacting Bose gases with a large number of particle species, as we will briefly discuss in Section \ref{sec:open_problems}.

When using expressions \eqref{relative pf-1}, \eqref{relative pf-2} to study the approach to the mean-field limit, $\nu\searrow 0$, we take $\lambda$ to be given by $\frac{\lambda_0 \nu^{2}}{N+1}$, with $\lambda_{0}$ kept constant, and choose the parameter $\rho$ in such a way that (up to a uniformly finite term) the phase $\theta(\sigma)$ defined in \eqref{phase} cancels the term \textit{linear} in $\sigma$ in 
\begin{equation}\label{resolvent id}
\int_{0}^{\infty} dt\, \mathrm{tr}\big[(K_{\Lambda}(\sigma)+t)^{-1} - (K_{\Lambda}(0)+t)^{-1} \big],
\end{equation}
which can be extracted by applying the second resolvent equation to $(K_{\Lambda}(\sigma)+t)^{-1}$. 
 
 Next, we derive an explicit expression for the resolvent of the operator $K_{\Lambda}(\sigma)$ introduced in \eqref{inv cov} on the space $L^{2}([0,\nu)\times \Lambda, d\tau\, dx)$, with periodic boundary conditions at $\tau=0, \nu$. We calculate the operator kernel of $\big(K_{\Lambda}(\sigma)+t\big)^{-1}$, i.e., the Green function of 
 $K_{\Lambda}(\sigma)$:
\begin{equation}\label{Green fu}
\big(K_{\Lambda}(\sigma)+t\big)^{-1}(\tau, x; \tau',x')
= \sum_{\ell=0}^{\infty} \Theta(\tau-\tau'+ \ell\nu)e^{(\tau'-\tau-\ell \nu)(\kappa_{0}+t)} \Gamma(\tau, \tau'-\ell\nu; i\sigma)_{xx'}\,,
\end{equation}
where $\tau$ and $\tau'$ belong to the interval $[0, \nu)$, $\Theta$ is the Heaviside step function, and 
$\Gamma(\tau, \tau';q)_{x,x'}$ is a \textit{heat kernel} solving the equation
\begin{equation}\label{heat kernel}
\frac{\partial}{\partial \tau} \Gamma(\tau, \tau';q)= \Big(\frac{\Delta}{2}-q(\tau, \cdot)\Big) \Gamma(\tau,\tau';q), \quad \text{ with }\,\, \Gamma(\tau,\tau;q) = {\bf{1}}\,,
\end{equation}
(and, as before, $\Delta$ is the Laplacian on $L^{2}(\Lambda, dx)$ with periodic boundary conditions at $\partial \Lambda$).
In Eq.~\eqref{Green fu}, the distribution $\sigma(\tau,x)\deq \sigma([\tau],x), [\tau]\deq \tau \text{ mod } \nu$, is defined to be periodic in $\tau$ with period $\nu$. 
Plugging \eqref{Green fu} into \eqref{relative pf-2}, taking the trace and integrating over $t$, we find that

\begin{equation}\label{relative pf-3}
\Xi_{\Lambda}^{\text{rel}}(\nu,\rho)=\int d\mu_{\lambda}(\sigma) \,e^{iN\theta(\sigma)}
 \exp\bigg\{N \sum_{\ell= 1}^{\infty} \frac{e^{-\ell \nu\kappa_{0}}}{\ell}\int_{\Lambda} du\big[\Gamma(\ell \nu,0; i\sigma)_{uu} -\Gamma(\ell \nu,0;0)_{uu}\big] \bigg\}\,,
 \end{equation}
 with $\theta(\sigma)\deq \frac{\rho}{\nu} \int_{0}^{\nu}d\tau\int_{\Lambda} dx\, \sigma(\tau,x)\,,$ as in \eqref{phase}. The series in the exponent of the integrand on the RHS of expression \eqref{relative pf-3} can be evaluated explicitly, and, using identity \eqref{det identity}, we find that
\begin{equation}\label{good formula}
\Xi_{\Lambda}^{\text{rel}}(\nu,\rho)=\int d\mu_{\lambda}(\sigma)\, e^{iN\theta(\sigma)} 
 \bigg[\frac{\det\big({\bf{1}}-e^{-\nu\kappa_0}\Gamma(\nu,0;i\sigma)\big)}{\det\big({\bf{1}}-e^{-\nu\kappa_0}\Gamma(\nu, 0;0))\big)}\bigg]^{N}\,.
\end{equation}
This is a wonderful, mathematically meaningful formula (which can also be derived directly from some of the formulae in Section \ref{sec:gc}, using a Hubbard-Stratonovich transformation; see \cite{FKSS-1}). Eq.~\eqref{good formula} is a convenient starting point to formally (and then rigorously) analyze the mean-field limit $\nu\searrow 0, \text{  with   }\lambda=\frac{\lambda_{0}\nu^{2}}{N+1}$; see \cite{FKSS-1}.

The RHS of expression \eqref{relative pf-3} can be analyzed with the help of the \textit{Feynman-Kac} formula for the heat kernel 
$\Gamma(\tau,\tau'; i\sigma)$, 
\begin{equation}\label{Feynman-Kac}
\Gamma(\tau,\tau';i\sigma)_{xy}= \int d\mathbb{W}^{\tau -\tau'}_{xy}(\omega)\, \exp\bigg\{-i\int_{0}^{\tau- \tau'} ds\, \sigma([s+\tau'], \omega(s))\bigg\}\,,
\end{equation}
where $d\mathbb{W}_{xy}^{\tau}(\omega)$ is the Wiener measure on Brownian paths starting at $x$ at time $0$ 
and reaching $y$ at time $\tau$. Thus,
\begin{equation}\label{FK}
\Gamma(\ell \nu,0; i\sigma)_{uu} -\Gamma(\ell \nu,0;0)_{uu}= \int d\mathbb{W}_{uu}^{\ell\nu}(\omega)
\big[e^{i\int_{0}^{\ell\nu}ds\,\sigma([s],\, \omega(s))} -1\big]\,.
\end{equation}
This identity \eqref{FK} shows that
\begin{equation}\label{negative Re}
\Re\Big(\Gamma(\ell \nu,0; i\sigma)_{uu} -\Gamma(\ell \nu,0;0)_{uu}\Big) \leq 0, \quad \forall \ell\,.
\end{equation}
Using also that $\vert e^{iN\theta(\sigma)}\vert =1$, we conclude that
\begin{equation}\label{stability}
\,\Xi_{\Lambda}(\nu,\rho) / \,\Xi^{(0)}_{\Lambda}(\nu) \leq 1\,,
\end{equation}
a bound that is already implicit in \eqref{gc density matrix}.

Reduced density matrices, $\gamma_{p}$, and imaginary-time Duhamel Green functions, $\mathcal{G}$, of the interacting gas can be calculated by integrating polynomials in the Green function 
$$K_{\Lambda}(\sigma)^{-1}(\tau, x; \tau',x')=\sum_{\ell =0}^{\infty} \Theta(\tau-\tau'+\ell \nu) \,
e^{-\kappa_{0}(\ell \nu + \tau- \tau')} \Gamma(\tau, \tau'-\ell \nu; i\sigma)_{xx'}$$ 
with the complex measure, $dP_{\Lambda}(\sigma)$, given by
\begin{equation}\label{int measure}
\Xi_{\Lambda}^{\text{rel}}(\nu,\rho)^{-1}\, \exp\bigg(iN\theta(\sigma) + N
\int_{0}^{\infty}dt\,\mathrm{tr} \big[(K_{\Lambda}(\sigma)+t)^{-1}- (K_{\Lambda}(0)+t)^{-1}\big]\bigg) d\mu_{\lambda}(\sigma)\,.
\end{equation}
As an example we give the expression for the 2-point Duhamel Green function
\begin{align}
\mathcal{G}(\tau, x, a;\tau',x',b) &\deq \nu^{-1}\mathrm{tr}\Big(\mathbb{P}_{\Lambda} e^{-\tau\mathbb{H}_{\nu,\Lambda}}\Phi^{*}_{\nu,a}(x) e^{(\tau-\tau')\mathbb{H}_{\nu,\Lambda}} \Phi_{\nu,b}(x')e^{\tau' \mathbb{H}_{\nu,\Lambda}}\Big)
\notag \\ \label{Duhamel}
&=\delta_{ab}\int dP_{\Lambda}(\sigma) K_{\Lambda}(\sigma)^{-1}(\tau, x; \tau',x'),
\end{align}
where $\mathbb{P}_{\Lambda}$ is the grand-canonical density matrix defined in Section \ref{sec:gc}, Eq.~\eqref{gc density matrix}.
Hence
\begin{equation}\label{RDM-2}
\gamma_{1}(x,a, x',b)\equiv \mathcal{G}(0,x,a;0,x',b)=\delta_{ab}\int dP_{\Lambda}(\sigma) K_{\Lambda}(\sigma)^{-1}(0, x; 0,x'),
\end{equation}
where $\gamma_{1}$ has been defined in \eqref{RDM}. We can then use identities \eqref{Green fu} and \eqref{Feynman-Kac} on the RHS of Eqs. \eqref{Duhamel} and \eqref{RDM-2} to express these quantities in terms of integrals over Brownian paths from $x$ to $x'$.

In the next section we use formulae \eqref{relative pf-3}, \eqref{Feynman-Kac}, \eqref{FK} and \eqref{Duhamel}, \eqref{RDM-2} to derive Ginibre's Brownian loop gas representation of interacting Bose gases \cite{Ginibre} (see also \cite{Su, Ue}) and Symanzik's loop gas representation of imaginary-time field theories \cite{Symanzik}, which emerges 
from Ginibre's representation in the mean-field limit, $\nu\searrow 0$.

\section{The loop gas representations of Ginibre and Symanzik and the mean-field limit of Bose gases} \label{sec:lg}
In order to introduce a gas of interacting Brownian paths and loops \textit{equivalent} to the interacting Bose gas, we expand the exponential in the integrand on the RHS of \eqref{relative pf-3}. We then use expression \eqref{FK} to rewrite the exponent appearing in the integrand on the RHS of \eqref{relative pf-3} and proceed to carry out the integration over $\sigma$, term by term. To describe the resulting expression, we define a \textit{``\,$2$-loop interaction''} potential
\begin{equation}\label{loop int}
V_{\nu}(\omega, \omega')\deq\frac{1}{2} \sum_{r=0}^{\ell(\omega)-1}\,\sum_{s=0}^{\ell(\omega')-1}\int_{0}^{\nu} dt\,
v(\omega(t+r\nu)-\omega'(t+s\nu))\,,
\end{equation}
We then find that
\begin{align}\label{loop gas}
\Xi_{\Lambda}^{\text{rel}}(\nu,\rho)= \text{const.}\sum_{n=0}^{\infty} \frac{N^{n}}{n!} &
\Biggl\{ \sum_{\ell_1,\dots,\ell_n =1}^{\infty} \int_{\Lambda} du_1\cdots \int_{\Lambda} du_n  {} \nonumber\\
& \times\Bigg[ \prod_{k=1}^{n} \frac{e^{-\ell_{k} \kappa(\rho) \nu}}{\ell_{k}}\int 
d\mathbb{W}_{u_{k} u_{k}}^{\ell_{k}\nu}(\omega_k) \Bigg]\, e^{-\sum_{i,j=1}^{n} 
\frac{\lambda}{\nu} V_{\nu}(\omega_{i}, \omega_{j})} \Biggr\},
\end{align}
with $\kappa(\rho)\deq\kappa_{0} - \lambda \rho \nu^{-2} \int dx\,v(x)$; see \eqref{quartic}. This is seen by using expression \eqref{Feynman-Kac} on the RHS of Eq.~\eqref{relative pf-3} and then carrying out the Gaussian integration over $\sigma$, using formula \eqref{Gauss int}, with $C$ as in \eqref{covariance}.

Similarly, we arrive at formulae for Duhamel Green functions and reduced density matrices in terms of a gas of interacting Brownian loops interacting with open Brownian paths that join arbitrary pairs of arguments of the Green functions corresponding to the same particle species.
For example,
\begin{multline}
\mathcal{G}(\tau,x,a;\tau',x',b) =  \delta_{ab}\, \Xi_{\Lambda}^{\text{rel}}(\nu,\rho)^{-1}\Bigg(\sum_{\ell_0=0}^{\infty} e^{-\kappa(\tau-\tau' + \ell_{0}\nu)}
 \int d\mathbb{W}_{xx'}^{\tau-\tau'+\ell_{0}\nu}(\omega_0) \Bigg) 
\\
\label{Duhamel-2}
\times
\Biggl\{\sum_{n=0}^{\infty} \frac{N^n}{n!} \Bigg[ \prod_{k=1}^{n}\sum_{\ell_k=1}^{\infty}  \frac{e^{-\kappa \ell_{k} \nu}}{\ell_{k}}\int_{\Lambda} du_{k} \int d\mathbb{W}_{u_{k} u_{k}}^{\ell_{k}\nu}(\omega_k) \Bigg]\, e^{-\sum_{i,j=0}^{n} \frac{\lambda}{\nu} V_{\nu}(\omega_{i}, \omega_{j})} \Biggr\}.
\end{multline}
Formulae \eqref{loop gas} and \eqref{Duhamel-2} constitute Ginibre's representation of an interacting Bose gas in the grand-canonical ensemble as a gas of interacting \textit{Brownian paths} and \textit{\mbox{-loops}}; see \cite{Ginibre}. We note that the number of particle species, $N$, appears as parameter on the RHS of \eqref{loop gas} and \eqref{Duhamel-2} that can be given arbitrary complex values.

Next, we derive Symanzik's representation \cite{Symanzik} of scalar imaginary-time field theories with quartic self-interactions -- see \eqref{class lim} and \eqref{Gibbs measure} -- as gases of interacting Brownian paths and -loops from Ginibre's representation. Our derivation is formal, but the resulting expressions are correct.\footnote{Readers concerned with mathematical rigor may want to introduce a lattice regularization of the expressions considered below and let the lattice spacing tend to $0$ at the end of the calculations; see \cite{FKSS-2}.} We return to expression \eqref{loop int} appearing in Eqs. \eqref{loop gas} and \eqref{Duhamel-2}. In these formulae we set 
\begin{equation}\label{coupling const}
\lambda= \frac{\lambda_0}{N+1}\cdot \nu^{2}\,.
\end{equation}
Inspecting expression \eqref{loop int} and plugging in \eqref{coupling const}, we see that the sums over 
$\ell_1,\dots,\ell_{n}$ on the RHS of \eqref{loop gas} can be viewed as \textit{Riemann sum approximations} to the quantity
\begin{align} \notag
Z_{\Lambda}= \underset{\delta\searrow 0}{\text{lim}} \,&\Bigg[\big(Z_{\Lambda}^{(0)}(\delta)\big)^{-N}\sum_{n=0}^{\infty}\frac{N^n}{n!}
\\
\label{field pf}
&\times\Bigg(\prod_{k=1}^{n} \int_{\delta}^{\infty} \frac{dT_k}{T_k}e^{-\kappa_{\delta}T_{k}} \int_{\Lambda} du_{k}\,\int 
 d\mathbb{W}_{u_k u_k}^{T_k}(\omega_k) \Bigg) \exp\bigg\{-\sum_{i,j=1}^{n} V_{0}(\omega_i, \omega_j) \bigg\}\Bigg]\,,
\end{align}
where $Z_{\Lambda}^{(0)}(\delta)$ is a $\delta$-dependent constant that is independent of $\lambda$ and $N$, the constant $\kappa_{\delta}$ (related to the chemical potential) is chosen so as to correctly implement Wick ordering in the limit $\delta\searrow 0$, and the \textit{``2-loop interaction''} potential $V_0$ is now given by 
\begin{equation}\label{Symanzik}
V_{0}(\omega, \omega') = \frac{1}{2} \int_{0}^{T} dt \int_{0}^{T'} dt' \,v(\omega(t)-\omega'(t'))\,.
\end{equation}
This is a variant of Symanzik's loop gas representation \cite{Symanzik} of imaginary-time scalar field theories.

\paragraph{Remark} The quantity in straight brackets, $[\cdot]$, on the RHS of \eqref{field pf} is regularized by imposing a strictly positive lower integration limit, $\delta>0$, on the integrations over the variables $T_k, k=1,\dots,n$; for, the function $T^{-1}$ is not integrable near $T=0$. After some further re-writing and appropriate normalization we will be able to pass to the limit 
$\delta\searrow 0$; see \eqref{Gaussian} and \eqref{pf}, below.

Let $d\bar{\mu}_{\lambda_0}(\eta)$ be the Gaussian probability measure on the space, $\mathcal{S}'(\Lambda)$, of tempered distributions on $\Lambda$ with mean 0 and covariance $\frac{\lambda_0}{N+1}v$; we should think of 
$\eta$ as being given by
\begin{equation}\label{eta}
\eta(x)\deq \nu^{-1}\int_{0}^{\nu} d\tau\, \sigma(\tau,x)\,,
\end{equation}
where the distribution of $\sigma$ is given by the Gaussian measure $d\mu_{\lambda}(\sigma)$, see \eqref{covariance}.
We then find that
\begin{align} \notag
\text{RHS of } \eqref{field pf} = \lim_{\delta\searrow 0} \,&\bigg[ \int d\bar{\mu}_{\lambda_0}(\eta)\,e^{i\theta_{\delta}(\eta)} 
\\
\label{inv Gaussian int}
&\times \exp\bigg\{N \int_{\delta}^{\infty} 
\frac{dT}{T} e^{-\kappa_{0} T} \int_{\Lambda} du\int d\mathbb{W}_{uu}^{T}(\omega) \Big(e^{i\int_{0}^{T} \eta(\omega(t))dt}-1\Big)\bigg\}\bigg],
\end{align}
where $\theta_{\delta}(\eta)\deq\theta_{\delta}\int_{\Lambda}dx\,\eta(x),$
 with $\theta_{\delta}$ a constant that diverges, 
as $\delta\searrow 0$, so as to implement the correct Wick ordering subtraction.

The Feynman-Kac formula says that
$$\int_{\Lambda}du \int  d\mathbb{W}_{uu}^{T}(\omega) \, e^{i\int_{0}^{T} \eta(\omega(t))dt} = \mathrm{tr}\big(e^{T(\frac{\Delta}{2}+i\eta)}\big)\,.$$
Using this identity on the RHS of \eqref{inv Gaussian int}, we obtain that
\begin{equation}\label{Gaussian}
\text{RHS of } \eqref{field pf}=\lim_{\delta \searrow 0} \int d\bar{\mu}_{\lambda_0}(\eta) \, e^{i\theta_{\delta}(\eta)} \exp\bigg\{N\int_{\delta}^{\infty} \frac{dT}{T} e^{-\kappa_{0}T} 
\Big[\tr\big(e^{T(\frac{\Delta}{2}+i\eta)}\big) - e^{T\frac{\Delta}{2}}\Big]\bigg\}\,.
\end{equation}
Carrying out the $T$-integration, choosing $\theta_{\delta}$ appropriately, and taking the limit 
$\delta \rightarrow 0$, we find that 
\begin{align}
Z_{\Lambda}&= \int d\bar{\mu}_{\lambda_0}(\eta) \, e^{iN\vartheta(\eta)} 
\exp\biggl\{-N\,\mathrm{tr}\bigg[\ln(-\frac{\Delta}{2} + \kappa_{0} - i\eta)-\ln(-\frac{\Delta}{2}+\kappa_{0})\bigg]_{\geq2}\biggr\} \nonumber \\ \label{pf}
&= \int d\bar{\mu}_{\lambda_0}(\eta)\, e^{iN\vartheta(\eta)} 
\bigg[\frac{\det(-\frac{\Delta}{2} + \kappa_{0} -i\eta)_{\text{ren}}}{\det(-\frac{\Delta}{2}+\kappa_{0})}\bigg]_{\text{ren}}^{-N}\,,
\end{align}
where $[(\cdot)]_{\geq 2}$ indicates that the term \textit{linear} in $\eta$ has been subtracted, the renormalized
quotient of determinants in the integrand on the RHS of \eqref{pf} is obtained by subtracting the term linear in $\eta$ in 
$$\ln \det\bigg[{\bf{1}} -i\biggl(-\frac{\Delta}{2}+\kappa_0\biggr)^{-1} \eta\bigg]\,,$$ and $\vartheta(\eta)= \vartheta \int_{\Lambda} dx\,\eta(x)\,, \vartheta<\infty,$ is a phase expressing a finite freedom in the choice of Wick ordering.
In one dimension and for lattice Bose gases,  these subtractions are superfluous; but, for continuum gases in two and three dimensions, they are essential  to arrive at a meaningful result, because the Wick ordering subtraction necessary to render the formal expression for the partition function $Z_{\Lambda}$ well defined is divergent. Applying identity \eqref{det identity} to the expression on the RHS of the first line in \eqref{pf} and using the second resolvent identity, we find that
\begin{equation}\label{pf-2}
Z_{\Lambda}= \int d\bar{\mu}_{\lambda_0}(\eta) \, e^{iN\vartheta(\eta)} e^{-N\mathcal{S}(\eta)}\,,
\end{equation}
where
\begin{equation}\label{action functional}
\mathcal{S}(\eta)\deq\int_{0}^{\infty} dt\,\mathrm{tr}\Bigg[\bigg(-\frac{\Delta}{2}+\kappa_{0}+t\bigg)^{-1}\,\eta\,\bigg(-\frac{\Delta}{2}+\kappa_{0}+t-i\eta\bigg)^{-1}\,\eta\,\bigg(-\frac{\Delta}{2}+\kappa_{0}+t\bigg)^{-1}\Bigg]\,.
\end{equation}
We observe that 
\begin{equation}\label{real part}
\Re\, \mathcal{S}(\eta) \geq 0\,.
\end{equation}
This is seen by noticing that the numerical range of the operator $\big(-\frac{\Delta}{2} + \kappa_{0}+t -i\eta\big)^{-1}$ appearing under the integral on the RHS of \eqref{action functional} is contained in $\lbrace z \mid \Re z \geq 0 \rbrace$.
 
By using the Hubbard-Stratonovich transformation \textit{in reverse} in Eq.~\eqref{pf}, we see that the partition function $Z_{\Lambda}$ is nothing but the partition function of the classical field theory introduced in Eqs.~\eqref{class lim} and \eqref{Gibbs measure}.

Although the arguments presented in this section are formal, it actually turns out that the expression for the relative partition function $\Xi_{\Lambda}^{\text{rel}}(\nu,\rho)$ given in \eqref{relative pf-2}, with \eqref{resolvent id}, converges to expression \eqref{pf-2} for the partition function of the classical field theory, as $\nu \searrow 0$ (i.e., in the mean-field limit). Furthermore, reduced density matrices of the Bose gases converge to correlation functions of the classical field theory. Proofs of all these results are given in \cite{FKSS-1}; (for an alternative approach see \cite{LNR}).

The idea underlying our proof of convergence to the mean-field limit presented in \cite{FKSS-1} is as follows. We return to expression \eqref{relative pf-2} for the relative partition function and then rewrite it as in \eqref{relative pf-3}. Using the important bounds \eqref{negative Re} and \eqref{real part} on the real parts of the exponents in the integrands of \eqref{relative pf-2} and \eqref{pf-2}, respectively, we conclude that it is enough to show that, after Wick subtraction,
the exponent in the integrand of formula \eqref{relative pf-2} (see also formula \eqref{good formula}), namely
$$\int_{0}^{\infty} dt \,\tr\Big[(K_{\Lambda}(0)+t)^{-1}\, \sigma\, (K_{\Lambda}(\sigma)+t)^{-1} \,\sigma \,(K_{\Lambda}(0)+t)^{-1}\Big],$$
converges in $L^{2}\big(\mathcal{S}'([0,\nu]\times \Lambda), d\mu_{\lambda_{0}\cdot \nu^{2}}(\sigma)\big)$ to
$$\int_{0}^{\infty} dt\,\mathrm{tr}\Bigg[\bigg(-\frac{\Delta}{2}+\kappa_{0}+t\bigg)^{-1}\,\eta\,\bigg(-\frac{\Delta}{2}+\kappa_{0}+t-i\eta\bigg)^{-1}\,\eta\,\bigg(-\frac{\Delta}{2}+\kappa_{0}+t\bigg)^{-1}\Bigg]\,,$$
using that when setting $\eta(x)\deq \nu^{-1}\int_{0}^{\nu} d\tau\, \sigma(\tau,x)$, see \eqref{eta},
integration over $d\bar{\mu}_{\lambda_0}(\eta)$ amounts to an integration over 
$d\mu_{\lambda_{0}\cdot \nu^{2}}(\sigma)$. Concrete convergence estimates are obtained by 
using \eqref{good formula}, \eqref{Feynman-Kac} and \eqref{FK} and carrying out the integration 
over $\sigma$. The details are somewhat tedious; see \cite{FKSS-1}.

\section{Further results and open problems} \label{sec:open_problems}

It is time to disclose what the representations derived in previous sections are good for, besides being neat and leading to a new (and rather transparent) proof of convergence of interacting Bose gases to their mean-field (classical field) limit, which we have sketched in the last section. 

\subsection{Classical particle limit}

We consider a Bose gas at positive temperature and density in the limit where the mass, $M$, of the particles in the gas tends to $\infty$, corresponding to the limit $\nu\searrow 0$, setting 
$$\lambda=\lambda_{0}= \text{const.}\,,$$
and choosing the chemical potential of the gas to depend on $M$ in a suitable manner. We claim that, in this limit, the partition function and the reduced density matrices of the Bose gas converge towards the partition function and the correlation functions of a classical gas of point particles with two-body interactions given by the potential $v(x)$.

To prove this claim, it is convenient to start from Ginibre's loop-gas representation \eqref{loop gas} of the grand partition function of the Bose gas, i.e., 
\begin{align}\notag
\Xi_{\Lambda}^{\text{rel}}\big(\nu, \rho(\nu)\big) = \text{const.}\sum_{n=0}^{\infty} \frac{N^n}{n!} &\Bigg[\prod_{k=1}^{n}\sum_{\ell_k=1}^{\infty}  \frac{e^{-\kappa(\nu) \ell_{k} \nu}}{\ell_{k}}\int_{\Lambda} du_{k} \int d\mathbb{W}_{u_{k} u_{k}}^{\ell_{k}\nu}(\omega_k) \Bigg]
\\
\label{Ginibre}
&\times \exp\bigg\{{-\sum_{i,j=0}^{n} \frac{\lambda}{\nu} V_{\nu}(\omega_{i}, \omega_{j})}\bigg\}\,,
\end{align}
with $\kappa(\nu)\deq\kappa_{0} - \lambda \rho(\nu) \nu^{-2} \int dx\,v(x)>0$ chosen in such a way that 
\begin{equation}\label{activity}
e^{-\kappa(\nu)\cdot \nu} \,\nu^{-\frac{d}{2}}\eqd z = \text{const.}, \quad \text{ for arbitrary }\, \nu>0\,.
\end{equation}
Using simple heat kernel estimates, one shows that, in the limit where $\nu\searrow 0$, only the terms with $\ell_{k}=1,
\forall\,k=1,\dots,n,$ survive, and, in view of expression \eqref{loop int} for $V(\omega_i,\omega_j)$, $ \Xi_{\Lambda}^{\text{rel}}\big(\nu, \rho(\nu)\big)$ is seen to converge to the \textit{grand partition function of the classical gas} given by
\begin{equation}\label{classical pf}
\Xi_{\Lambda}(z)\deq \sum_{n=0}^{\infty} \frac{(zN)^{n}}{n!} \Bigg[\prod_{k=1}^{n} \int_{\Lambda} du_{k}\,\Bigg]
\exp\bigg\{-\frac{\lambda_0}{2}\sum_{i,j=1}^{n} v(u_{i}-u_{j}) \bigg\}\,.
\end{equation}

\subsection{Thermodynamic limit}
\label{Thermodynamic limit}
This paragraph is expository, and the results described here are \textit{not} new; see \cite{Ginibre, Brydges, Ueltschi, Fernandez} and references given there, and \cite{FKSS-2} for a detailed review. We therefore just describe a few \textit{basic ideas} underlying the analysis of the thermodynamic limit of high-temperature, low-density quantum and classical gases. Our goal is to prove convergence of the Gibbs potential (per unit volume),
$$\omega_{\Lambda}(\nu, \rho)\deq\frac{1}{\vert \Lambda \vert} \ln \Xi_{\Lambda}(\nu, \rho)\,,$$
and of the reduced density matrices, $\gamma_{p}$, as $\Lambda \nearrow \mathbb{R}^{d}$, and to establish analyticity properties of the limiting expressions in the inverse temperature and the chemical potential, for large enough values of the temperature and suffciently low densities. Instead of re-introducing temperature and chemical potential, we vary the coupling constant 
$\lambda_0$ and the parameter $\kappa_0$ related to the chemical potential. We propose to use \textit{cluster expansions} \cite{Brydges, Ueltschi, Fernandez} to analyze the approach to the thermodynamic limit. Sufficient conditions for the convergence of these expansions are that (i) the two-body potential $v(x)$ is non-negative (``repulsive'') and decays sufficiently rapidly (more precisely integrably fast), as $\vert x \vert \rightarrow \infty$, (ii) the coupling constant $\lambda_{0}$ is small enough, with 
$\Re \lambda_0 \geq 0$, and (iii) the density of the gas is small enough (e.g., $\rho=0$ and $\kappa_{0}\not= 0$, with 
$\Re \kappa_0 >0$ large enough, depending on $\lambda_0$ and $\nu$, in Eqs. \eqref{quartic} and \eqref{path integral}; $\vert z \vert$ small enough in Eq.~\eqref{classical pf}).

We focus our attention on the cluster expansion of the grand partition function of the quantum gas. To begin with, we set
\begin{equation}\label{G fu}
\exp\bigg\{-\frac{\lambda_0}{\nu}V(\omega, \omega')\bigg\}\eqd {\bf{1}}+G_{\lambda_0}(\omega, \omega')
\end{equation}
and note that
$$\vert G_{\lambda_0}(\omega, \omega')\vert=\mathcal{O}(\lambda_0)$$
can be made arbitrarily small, uniformly in the loops $\omega$ and $\omega'$, by choosing $\lambda_0$ sufficiently small, and exhibits integrable fall-off in dist$(\omega, \omega')$. The cluster expansion -- in the present context called \textit{Mayer expansion} -- is obtained by expanding the products
$$\prod_{1\leq i<j\leq n} \Big[{\bf{1}}+G_{\lambda_0}(\omega_i, \omega_j)\Big]\,, \quad n=1,2,3, \dots\,,$$
in the integrands of the terms appearing in, for example, the grand partition function in powers of the small quantities $G_{\lambda_0}(\omega_i, \omega_j), i \not= j$. One then applies the \textit{linked cluster theorem} 
to the resulting expression; (i.e., one considers the logarithm of the grand partition function). This results in a sum of terms that can be labelled by \textit{connected diagrams}: One assigns a \textit{vertex} of the diagram to each loop, 
$\omega_i$, appearing in a term and assigns a \textit{line} (edge) to every pair, \mbox{$(\omega_i, \omega_j)$,} of loops appearing in a factor 
\mbox{$G_{\lambda_0}(\omega_i, \omega_j)$} of the term. To prove \textit{convergence of the Mayer expansion}, one fixes a spanning tree in every connected diagram labelling a term in the expansion. One then sums over all contributions corresponding to diagrams with the same spanning tree. In the resulting expressions, one estimates the factors 
\mbox{exp$\big\{-\frac{\lambda_0}{\nu}V(\omega_i, \omega_j)\big\}$} from above by ${\bf{1}}$, 
(keeping $\Re\lambda_0$ non-negative). Standard estimates and combinatorics then enable one to 
prove convergence, assuming that 
$\vert \lambda_0 \vert$ is small enough, with 
$\Re \lambda_0$ non-negative, and $\Re \kappa_0>0$ large enough (depending on $\lambda_0$).

The analysis of classical gases of point particles with non-negative integrable two-body potentials is simpler; see, e.g., \cite{Fernandez} and references given there.

\subsection{Bose-Einstein condensation in the $N\rightarrow \infty$ limit}
We are interested in studying the behavior of a Bose gas when the number, $N$, of particle species is very large, choosing 
$\lambda=\frac{\lambda_0\nu^{2}}{N+1}$. Inspecting formulae \eqref{relative pf-1} and \eqref{int measure}, \eqref{Duhamel}, we observe that one can apply the saddle-point method to study the large-$N$ asymptotics of the (relative) grand partition function $\Xi_{\Lambda}^{\text{rel}}(\nu,\rho)$ given in \eqref{relative pf-1} and the convergence of the 2-point Duhamel Green function given in \eqref{Duhamel}, as $N\rightarrow \infty$. In fact
\begin{equation*}
\mathcal{G}(\tau, x, a;\tau',x',b) \underset{N\rightarrow \infty}{\longrightarrow} \delta_{ab}K_{\Lambda}(\sigma_{*})^{-1}(\tau,x;\tau',x')\,,
\end{equation*}
where $\sigma_{*}$ is the critical point of the functional 
\begin{equation}\label{saddle point}
\mathfrak{S}(\sigma)\deq \frac{1}{2} \langle\sigma, C^{-1} \sigma\rangle_{L^{2}} - i\theta(\sigma)
-\int_{0}^{\infty}dt\,\tr\big[(K_{\Lambda}(\sigma)+t)^{-1} - (K_{\Lambda}(0)+t)^{-1}\big]\,,
\end{equation}
with $C$ the covariance of the Gaussian measure $d\mu_{\lambda}(\sigma)$ given in \eqref{covariance}. This functional appears in the exponent of the measure $dP_{\Lambda}(\sigma)$ defined in \eqref{int measure}. Using Eq.~\eqref{good formula}, the last term on the RHS can be re-written as
$$\tr\Big\{\ln\big[{\bf{1}}-e^{-\nu\kappa_0}\Gamma(\nu,0;i\sigma)\big]- \ln\big[{\bf{1}}-e^{-\nu\kappa_0}\Gamma(\nu,0;0)\big]\Big\}$$
The critical point of the functional $\mathfrak{S}(\sigma)$ can be determined on the basis of calculations similar to (but slightly more complicated than) those used in \cite[Section 3.3]{DI} to determine the large-$N$ asymptotics of the \textit{$N$-vector model}; see also \cite{MZ}.

Inspecting expression \eqref{inv cov} for $K_{\Lambda}(\sigma)$, we find that the 2-point Duhamel Green function approaches the 2-point Duhamel Green function of an ideal Bose gas and, similarly, the reduced density matrix 
$\gamma_1$ converges to the reduced density matrix of an ideal Bose gas
with a \textit{renormalized chemical potential}, 
\mbox{$\kappa_{0} \mapsto \kappa_{0} + i\sigma_{*}\equiv \kappa_{\text{ren}}$.}

The same conclusion can be reached by analyzing the behavior of terms in the cluster expansion (see 
Section \ref{Thermodynamic limit}, above) of $\mathcal{G}(\tau, x, a;\tau',x',b)$, as $N\rightarrow \infty$: 
Every term in this expansion displays a path, 
$\omega$, starting at a point $x\in \Lambda$ at time $\tau$ and reaching $x'\in \Lambda$ at a time $\tau'+\ell \nu$, for some $\ell=1,2,3,\dots$. This path interacts with loops, $\omega'$, through factors 
$G_{\lambda_0}(\omega, \omega')$; see \eqref{G fu}. Using the 
diagrammatic rules introduced in Section \ref{Thermodynamic limit}, we find that, in the limit $N\rightarrow \infty$, only diagrams contribute where \textit{trees} are attached to $\omega$. The reason is that every vertex in diagrams labelling terms contributing to 
$\mathcal{G}$, which corresponds to some Brownian loop traversed by $N$ species of particles, is proportional to $N$, while every line is proportional to $\frac{1}{N}$, thanks to our choice, $\lambda=\frac{\lambda_0\nu^{2}}{N+1}$, of the coupling constant. The contributions corresponding to those tree diagrams can be resummed and are seen to merely renormalize the death rate of the path $\omega$, i.e., the chemical potential of the gas.

We conclude that, in the limit where $N\rightarrow \infty$, an interacting Bose gas of $N$ species of particles exhibits the onset of \textit{Bose-Einstein condensation}, as $\kappa_{0}$ approaches a critical value, $\kappa_{*}$, that depends 
on the strength of the coupling constant $\lambda_0$.
This conclusion is well known for the \textit{N-vector model} introduced in Eqs. \eqref{class lim} and \eqref{Gibbs measure} describing the mean-field limit of Bose gases: The $N\rightarrow \infty$ limit of this model is equivalent to the so-called \textit{spherical model}, which was introduced and shown to exhibit a phase transition by Berlin and Kac in \cite{B-K}.

What about large, but finite values of $N$? To answer this question one might attempt to make use of \textit{renormalization group methods} to study the functional integrals appearing on the right sides of Eqs. \eqref{relative pf-1}, \eqref{good formula} and \eqref{Duhamel}, taking advantage of the fact that, for large values of $N$, one can use an expansion of the functional $\mathfrak{S}(\sigma)$ introduced in Eq.~\eqref{saddle point} around the critical point $\sigma_{*}$. For large values of $N$, the constant term, $\mathfrak{S}(\sigma_{*})$, and the quadratic terms (the Hessian of $\mathfrak{S}$ at the point $\sigma_{*}$) dominate in this expansion; higher-order corrections are proportional to inverse powers of $N$ (weak-coupling regime). This ought to enable one to analyze the properties of the functional measure $dP_{\Lambda}(\sigma)$, for sufficiently large values of $N$. (A possible alternative method would consist in generalizing the so-called \textit{lace expansion} introduced in \cite{BS}; but it is far from clear whether this will work.) 

Our discussion leads us to expect that one should be able to prove rigorously that an interacting Bose gas of the type studied in this paper exhibits Bose-Einstein condensation accompanied by the spontaneous breaking of the gauge symmetry (of the first\, kind), $\varphi_{a} \mapsto e^{i\theta} \varphi_{a}, a=1,\dots,N$, in 
$d\geq 4 \text{ dimensions, and for sufficiently large values of } N.$

\subsection{The limit $N\rightarrow 0$ and the SAW conjecture}
It is interesting to study the behavior of the 2-point Duhamel Green function and the reduced density matrices using the diagrammatic analysis sketched in Section \ref{Thermodynamic limit}, above. We then find that, in the limit where $N\searrow 0$, with 
$\lambda= \lambda_{0}\nu^{2}$ independent of $N$, every vertex corresponding to a Brownian loop in a diagram gives rise to a factor $N$ in the contribution labelled by that diagram. 
Thus, only terms labelled by diagrams \textit{without} any vertices associated with Brownian loops contribute to a 2-point Duhamel Green function or a reduced density matrix $\gamma_1$, as $N\searrow 0$. 
The model resulting in this limit turns out to be a \textit{regularized version} of the Edwards model \cite{E,Westwater} of weakly self-avoiding walks; (see \cite{FKSS-2}). Our findings represent a (fairly obvious) generalization of de Gennes' analysis \cite{de Gennes} of polymer chains modelled as (weakly) self-avoiding walks. 
We expect that, for sufficiently small \textit{positive} values of 
$N\in \mathbb{R}$ and in dimension $d\geq 4$, these models have a critical point in the same universality class as the ideal Bose gas and the usual model of standard random walks. If we were asked to propose a strategy to prove this rigorously we would presumably recommend to apply the \textit{supersymmetry methods} developed in \cite{BBS-1}; see also \cite{BBS-2}.

\subsection{BCS theory of superconductivity}
The analysis of BCS superconductivity in interacting electron gases with attractive forces in the Cooper channel leads one to study certain imaginary-time models of a self-interacting complex scalar field; see, e.g., \cite{AGD, NJL}, and \cite{CFS} and references given there. The expressions turning up in these models are reminiscent of those encountered in Sections \ref{sec:pi} and \ref{sec:lg} in our study of interacting Bose gases and of their mean-field limits. One expects that rather similar analytical methods can be applied in both cases to study the expected phase transition accompanied by spontaneous breaking of the $U(1)$-gauge symmetry of the first kind.

 \subsection{Triviality of $\lambda \vert {\phi} \vert^{4}_{d}$-theory in dimension $d \geq 4$}
 It is known that, in dimensions $d\geq 4$, the limit, as $v(x) \rightarrow \delta^{(d)}(x)$, of a Bose gas with a repulsive two-body potential $v>0$ is an \textit{ideal gas} without any interactions. In other words, in four or more dimensions,  Bose gases of particles with a point-like hard core of arbitrary strength 
are identical to non-interacting gases. Intuitively, this can be understood to be a consequence of the 
fact that, for $d\geq 4$, two Brownian paths starting at different points in physical space 
$\mathbb{R}^{d}$ never intersect (see \cite{ET}), so that a 
$\delta$-function repulsive two-body potential does not have any effect. In contrast, in one, two and three dimensions, Bose gases with repulsive $\delta$-function potentials exhibit non-trivial interaction effects. It is expected that, for $d=1,2$ and 3, one can rigorously prove, without any heroic efforts, that these gases converge towards Euclidean 
$\lambda\vert \boldsymbol{\phi} \vert^{4}_{d}$-theories 
 in the mean-field limit, as discussed in Section \ref{sec:lg}; see \cite{FKSS-1, FKSS-2}. Some techniques expected to be relevant to supply proofs have been developed in \cite{RW}).

 Thus, a Bose gas with repulsive $\delta$-function two-body potentials in $d$-dimensional space $\mathbb{R}^{d}$ can be viewed as a regularization of the Euclidean (imaginary-time) 
 $\lambda\vert \boldsymbol{\phi} \vert^{4}_{d}$-theory of a complex scalar field $\boldsymbol{\phi}$ in $d$ dimensions. Since the Bose gas with repulsive $\delta$-function two-body potentials is non-interacting in four or more dimensions, this leads to the highly plausible \textit{conjecture} that the same is true for its mean-field limit, namely that, for $d\geq 4$, the 
 $\lambda\vert \boldsymbol{\phi} \vert^{4}_{d}$-theory of a complex scalar field $\boldsymbol{\phi}$ is equivalent to a Gaussian (free) field theory.
 The same conclusion is expected to be true for an arbitrary number, $N$, of species of particles in the gas and $N$-component complex scalar fields. Partial results in this direction have been proven in \cite{JF}, 
 and beautiful results for the Ising model and $\phi^{4}$-theory of a real one-component scalar field, $\phi$,
 including the four-dimensional theory, have been established in \cite{A-DC} based on methods developed in \cite{Aizenman}. 
 
More details on the results discussed in Section \ref{sec:lg} and in this section and proofs can be found in \cite{FKSS-1} and \cite{FKSS-2}.

 \section{Epilogue: It is not only science that matters} \label{sec:ep}
Joel Lebowitz is a uniquely charismatic colleague. He has made essential contributions to a very large number of interesting and deep results in general statistical physics. Through his organization of \textit{one hundred and twenty three} conferences on statistical mechanics, which first took place at Yeshiva University and later at Rutgers, and in many other ways that include his dedicated work over many years as the chief editor of this journal, he has done an invaluable service to the community of people interested in or working on problems in statistical physics and to the sociology of this community. 
 
However, what is equally or even more admirable is Joel's \textit{commitment to important good causes}, in particular to the promotion of human rights and to the support of colleagues in many countries who were or are deprived of basic human rights.
In these rather precarious times, the efforts of people like Joel are essential, even if they often take a long while to bear fruit.
We very much hope that the example Joel has set in science and his outstanding human qualities will continue to inspire younger generations to follow his lead!
 
 \vspace{1em}
 \begin{center}
 Thank you Joel!
 \end{center}
 \vspace{1em} 
 \begin{center}
\textit{R\'{e}veillez-vous, indignez-vous, engagez-vous!}
\\[0.5em]
\hspace{5cm} --- S.\ Hessel
 \end{center}

\bigskip

\begin{center}
-----
\end{center}

\bigskip

\noindent
J\"urg Fr\"ohlich, ETH Z\"urich, Institute for Theoretical Physics, \href{mailto:juerg@phys.ethz.ch}{juerg@phys.ethz.ch}.
\\[0.3em]
Antti Knowles, University of Geneva, Section of Mathematics, \href{mailto:antti.knowles@unige.ch}{antti.knowles@unige.ch}.
\\[0.3em]
Benjamin Schlein, University of Z\"urich, Institute of Mathematics, \href{mailto:benjamin.schlein@math.uzh.ch}{benjamin.schlein@math.uzh.ch}.
\\[0.3em]
Vedran Sohinger, University of Warwick, Mathematics Institute, \href{mailto:V.Sohinger@warwick.ac.uk}{V.Sohinger@warwick.ac.uk}.


\begin{thebibliography}{References}

\bibitem{FKSS-1} J. Fr\"{o}hlich, A. Knowles, B. Schlein and V. Sohinger, \textit{The mean-field limit of quantum Bose gases at positive temperature}, preprint arXiv: 2001.01546v1

\bibitem{FKSS-2} J. Fr\"{o}hlich, A. Knowles, B. Schlein and V. Sohinger, \textit{Interacting loop ensembles and Bose gases}, in preparation

\bibitem{B-K} T. H. Berlin and M. Kac, \textit{The Spherical Model of a Ferromagnet}, Phys. Rev. {\bf{86}}, 821- 835 (1952)

\bibitem{DI} J.-M. Drouffe and C. Itzykson, \textit{Statistical field theory}, vol. 1, Cambridge University Press, Cambridge \& New York, 1989

\bibitem{MZ} M. Moshe and J. Zinn-Justin, \textit{Quantum field theory in the large N limit: a review}, Physics Reports {\bf{385}}, 69-228 (2003)

\bibitem{CFS} T. Chen, J. Fr\"{o}hlich and M. Seifert, \textit{Renormalization Group Methods: Landau-Fermi Liquid and BCS Superconductor}, in: ``Fluctuating Geometries in Statistical Mechanics and Field Theory'', proceedings of Les Houches 62, 
F. David, P. Ginsparg and J. Zinn-Justin (eds.), Elsevier Science, Amsterdam 1995

\bibitem{A} J. O. Andersen, \textit{Theory of weakly interacting Bose gases}, Rev. Mod. Phys. {\bf{76}} 599-639 (2004)

\bibitem{LNR} M. Lewin, P.T. Nam and N. Rougerie, \textit{Classical Field Theory Limit of Many-Body Quantum Gibbs States in 2D and 3D}, arXiv:1810.08370, to be published

\bibitem{Ginibre} J. Ginibre, \textit{Some applications of functional integration in statistical mechanics}, in: ``M\'{e}canique quantique et th\'{e}orie quantique des champs, proceedings of Les Houches, C. De Witt and R. Stora (eds.), 1970

\bibitem{Su} A. S\"{u}to, \textit{Percolation transition in the Bose gas}, J. Phys. {\bf{A 26}}, 4689–4710 (1993);
\textit{Percolation transition in the Bose gas II}, J. Phys. {\bf{A 35}}, 6995–7002 (2002)

\bibitem{Ue} D. Ueltschi, \textit{Feynman cycles in the Bose gas}, J. Math. Phys. {\bf{47}}, 123303-1-15 (2006)

\bibitem{de Gennes} P. G. de Gennes, \textit{Exponents for the excluded volume problem as derived by the Wilson method}, Phys. Lett. {\bf{38A}}, 339-340 (1972)

\bibitem{DP}B. Duplantier and P. Pfeuty, \textit{$O(n)$ field theory with $n$ continuous as a model for equilibrium polymerisation}, J. Physics A: Math and General, {\bf{15}}, (1982); and references given there

\bibitem{Aizenman} M. Aizenman, \textit{Geometric analysis of $\phi^{4}$ fields and Ising models , I, II,}, Commun. Math. Phys. {\bf{86}} (1), 1-48 (1982)

\bibitem{JF} J. Fr\"{o}hlich. \textit{On the triviality of $\lambda\phi^{4}_{d}$ theories and the approach to the critical point in $d \underset{(=)}{>} 4$ dimensions}, Nuclear Physics B {\bf{200}} (2), 281-296 (1982)

\bibitem{A-DC} M. Aizenman and H. Duminil-Copin, \textit{Marginal triviality of the scaling limits of critical 4D Ising and $\phi^{4}_{4}$ models}, preprint, arXiv:1912.07973v1

\bibitem{BFKT} T. Balaban, J. Feldman, H. Kn\"{o}rrer and E. Trubowitz, \textit{A functional integral representation for many boson systems. I: The partition function}, Ann. Henri Poincar\'{e} {\bf{9}}, 1229–1273 (2008); \textit{A functional integral representation for many boson systems. II: Correlation functions}, Ann. Henri Poincar\'{e} {\bf{9}}, 1275-1307 (2008)

\bibitem{BS} D. Brydges and T. C. Spencer, \textit{Self-avoidiung walk in $5$ or more dimensions}, Commun. Math. Phys. {\bf{97}}, 125-148 (1985)  

\bibitem{BBS-1} R. Bauerschmidt, D. Brydges and G. Slade, \textit{Critical Two-Point Function of the $4$-Dimensional Weakly Self-Avoiding Walk}, Commun. Math. Phys. {\bf{338}}, 169-193 (2015)

\bibitem{BBS-2} R. Bauerschmidt, D. Brydges and G. Slade, \textit{Logarithmic correction for the susceptibility of the $4$-dimensional weakly self-avoiding walk: a renormalisation group analysis}, Commun. Math. Phys. {\bf{337}}, 817-877 (2015); \textit{Scaling limits and critical behaviour of the $4$-dimensional $n$-component $\vert \varphi \vert^{4}$ spin model}, J. Stat. Phys. {\bf{157}}, 692-742 (2014)

\bibitem{Ruelle-1} D. Ruelle, ``Statistical Mechanics -- Rigorous Results'', World Scientific, Imperial College Press, London 1999; ($1^{st}$ edition published in 1969 by W. A. Benjamin Inc.)

\bibitem{Ruelle-2} D. Ruelle, \textit{Analyticity of Green's functions of dilute quantum gases}, J. Math.  Phys.{\bf{12}},(1971) 901-903 (1971); (see also: J. Fr\"{o}hlich, \textit{The reconstruction of quantum  fields from Euclidean Green's functions at arbitrary temperatures}, Helv. Phys. Acta {\bf{48}}, 355-369 (1975))

\bibitem{LNR-2} M. Lewin, P. T. Nam and N. Rougerie, \textit{Derivation of nonlinear Gibbs measures from many-body quantum mechanics}, J. de l'\'{E}cole Polytechnique - Math\'{e}matiques {\bf{2}}, 65-115 (2015)

\bibitem{LNR-3} M. Lewin, P. T. Nam, N. Rougerie, \textit{Gibbs measures based on 1D (an)harmonic oscillators as mean-field limits}, J. Math. Phys. \textbf{59}, no.4, 041901 (2018).
\bibitem{LNR-4}  M. Lewin, P.T. Nam, N. Rougerie, \textit{Classical field theory limit of 2D many-body quantum Gibbs states}, preprint, arXiv: 1805.08370v3.

\bibitem{FKSS-3} J. Fr\"{o}hlich, A. Knowles, B. Schlein and V. Sohinger, \textit{Gibbs measures of nonlinear Schr\"{o}dinger equations as limits of many-body quantum states in dimensions $d\leq 3$}, Commun. Math. Phys. {\bf{356}}, 883-980 (2017)

\bibitem{FKSS-4} J. Fr\"{o}hlich, A. Knowles, B. Schlein and V. Sohinger, \textit{A microscopic derivation of time-dependent correlation functions of the 1D cubic nonlinear Schr\"{o}dinger equation}, Adv. Math. {\bf{353}}, 67-115 (2019)

\bibitem{Sohinger_2019} V.Sohinger, \textit{A microscopic derivation of Gibbs measures for nonlinear Schr\"{o}dinger equations with unbounded interaction potentials}, preprint, arXiv:1904.08137v2 

\bibitem{Pizzo} A. Pizzo, \textit{Bose particles in a box I-III}, Preprints 2015

\bibitem{GJ} J. Glimm and A. Jaffe, \textit{Quantum Physics -- a Functional Integral Point of View}, Springer-Verlag, New York, Berlin, Heidelberg 1987

\bibitem{Simon} B. Simon, \textit{The $P(\phi)_{2}$ Euclidean (Quantum) Field Theory}, Princeton University Press, Princeton NJ 1974

\bibitem{Bourgain-1} J. Bourgain, \textit{Periodic nonlinear Schr\"{o}dinger equation and invariant measures}, Commun. Math. Phys. {\bf{166}}, 1-26 (1994)

\bibitem{Bourgain-2} J. Bourgain, \textit{Invariant measures for the 2D-defocusing nonlinear Schr\"{o}dinger equation}, Commun. Math. Phys. {\bf{176}}, 421-445 (1996)

\bibitem{Bourgain-3} J. Bourgain, \textit{Invariant measures for the Gross-Pitaevskii equation}, J. Math. Pures 
Appl. {\bf 76}, 649-702 (1997)

\bibitem{Dirac} P. A. M. Dirac, \textit{The Lagrangian in quantum mechanics}, Phys. Zeitschrift der Sowjetunion {\bf{3}}, 64-72 (1933)

\bibitem{Symanzik} K. Symanzik, \textit{Euclidean quantum field theory}, in: \textit{Rendiconti della Scuola Internationale di Fisica ``Enrico Fermi''}, XLV Corso, ``Teoria quantistica locale'',  R. Jost (ed.), Academic Press, New York, 1969

\bibitem{Brydges} D. C. Brydges, \textit{A short course in cluster expansions}, in:  Proceedings of the 1984 Les Houches School on ``Critical Phenomena, Random Systems, Gauge Theories'', K. Osterwalder and R. Stora (eds.), pp 129-183, Elsevier, 1984

\bibitem{Ueltschi} D. Ueltschi, \textit{Cluster expansions and correlation functions}, Mosc. Math. J. {\bf{4}}, 511–522 (2004).

\bibitem{Fernandez} R. Fernandez and A. Procacci, \textit{Cluster expansion for abstract polymer models -- New bounds from an old approach}, Commun. Math. Phys. {\bf{274}}, 123–140 (2007);\\
R. Fernandez and N. T. Xuan, \textit{Convergence of cluster and virial expansions for repulsive classical gases}, 
preprint, arXiv:1909.13257v1.


\bibitem{E} S. F. Edwards, \textit{The statistical mechanics of polymers with excluded volume}, Proc. Phys. Soc. London {\bf{85}}, 613-624 (1965) 

\bibitem{Westwater} J. Westwater, \textit{On Edwards' Model for Polymer Chains, I}, Commun. Math. Phys. {\bf{72}}, 131-174 (1980)

\bibitem{ET} P. Erd\H{o}s and S. J. Taylor, \textit{Some problems concerning the structure of random walk paths}, Acta Math. Acad. Sci. Hung. {\bf{11}}, 137-162 (1960); and \textit{Some intersection properties of random walk paths}, Acta Math. Acad. Sci. Hung. {\bf{11}}, 231-248 (1960)

\bibitem{RW} V. Rivasseau and Z. Wang, \textit{Constructive Renormalization for $\Phi^{4}_{2}$ Theory with Loop Vertex Expansion}, J. Math. Phys. {\bf{53}}, 042302 (2012); and \textit{Corrected Loop Vertex Expansion for 
$\Phi^{4}_{2}$ Theory}, J. Math. Phys. {\bf{56}}(6), 062301 (2015)

\bibitem{AGD} A. A. Abrikosov, L. P. Gorkov and I. E. Dzyaloshinsky, \textit{Methods of Quantum Field Theory in Statistical Physics}, Dover Publ., New York 1975

\bibitem{NJL} Y. Nambu and G. Jona-Lasinio, \textit{Dynamical Model of Elementary Particles Based on an Analogy with Superconductivity. I}, Phys. Rev. {\bf{122}}, 345-358 (1961)

\end{thebibliography}
\end{document}